Title:    Shrinkage of cane (*Arundo donax* L.) II

Effect of drying condition on the intensity of cell collapse

Authors:    Eiichi OBATAYA (✉)

Institute of Wood Technology, Akita Prefectural University
016-0876 Akita, Japan
Phone : 0185-52-6984, Fax : 0185-52-6976
obataya@iwt.akita-pu.ac.jp

Joseph GRIL

Laboratoire de Mécanique et Génie Civil, Université Montpellier 2
860 Rue St.Priest, 34000 Montpellier, France
Phone : +33 4 67 14 34 33,    Fax : +33 4 67 14 47 92
jgril@lmgc.univ-montp2.fr

Patrick PERRÉ

Laboratoire de Recherche sur le Matériau Bois
UMR INRA/ENGREF/UHP 1093
ENGREF 14, Rue Girardet, F-54042 Nancy, France
Phone : +33 3 83 39 68 00,    Fax : +33 3 83 30 22 54 (Fax)
perre@engref.fr

**ABSTRACT**

To improve the drying method in the manufacture of woodwind reeds, green canes (*Arundo donax* L.) were dried under various humidity-temperature conditions and the intensity of cell collapse was evaluated from the swelling due to steaming involving the recovery of collapse. At 30°C, the intensity of collapse was increased by slower drying. It was considered that: 1) slower drying resulted in higher sample temperature in the early stage of drying to increase the collapse; 2) rapid drying stiffened the surface of sample and such "shell" prohibited the following collapse; 3) slower drying *i.e.* longer loading of liquid tension caused more remarkable and/or frequent viscoelastic yields of cells. Consequently the intensity of collapse increased when the cane was dried from its waxy outer surface or in the presence of node: both of them retarded the drying. On the other hand, higher drying temperature caused greater intensity of collapse in spite of faster drying. It was suggested that the thermal softening of cane cells leads to easier yield of the cell wall, at the same time the rapid drying does not allow the recovery of collapse after the disappearance of free water. These results indicated that faster drying at lower temperature is preferable for drying cane with less collapse.

## INTRODUCTION

A cane (*Arundo donax* L.) is widely used for the vibrating plate (reed) of woodwind instruments such as clarinet and saxophone. In the previous paper, we exhibited serious collapse of parenchyma cells during drying.[1] Since the recovery of collapse sometimes causes problematic swelling of reed, it is necessary to establish a drying method involving less collapse. However, the mechanism of cane collapse is still unclear whereas that of wood has so far been discussed in detail. The cane for woodwind reeds is usually harvested in winter and dried in open air without being separated into individual internodes. In this case, the cane dries very slowly in the presence of node obstructing the dispersion of water along its longitudinal direction. In general, slower drying is recommended for wood to reduce the risk of check and split due to steep moisture gradient. However, it has been suggested that slower drying of wood and bamboo resulted in greater intensity of collapse.[2,3] If the slower drying of cane also induces more remarkable collapse, we will have to reconsider the conventional drying method so far employed by many reed manufacturers. In this paper, we describe the effects of drying condition on the intensity of collapse to suggest a better drying method with less collapse.

### Materials and Method

**Cane sample**

Two green canes (2 years old) were obtained at a farm of Marca Reed Inc. These canes were separated into several poles and wrapped with poly-vinylidene chloride film to prevent the dehydration until the experiments. The position of internode was identified by a node number (Nr) from 1 (bottom) to 24 (top). To obtain plural specimens from the same internode, short and/or twisted internodes were excluded. Figure 1 shows the average external diameter and thickness of internodes tested.

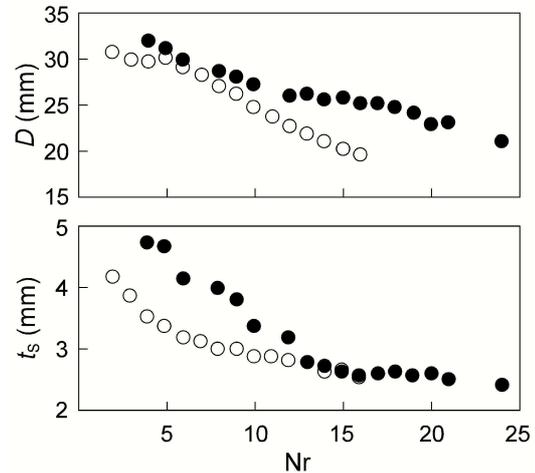

Fig.1. Average diameter ($D$) and thickness ($t_S$) of cane internodes tested plotted against the node number (Nr). *Open circles*, a cane pole used for testing the effect of node; *Closed circles*, for testing the effects of drying conditions. The $D$ was measured in green state and the $t_S$ was measured after steaming.

**Drying of cane specimens**

One cane pole was divided into short tubes as shown in Fig.2a. The node was removed from 7 internodes (Nr=3, 5, 7, 9, 11, 13 and 15) while it remained in the other 8 internodes (Nr=2, 4, 6, 8, 10, 12 and 16). The tubes were then dried at 20°C and 65% relative humidity (*RH*) without the circulation of air.

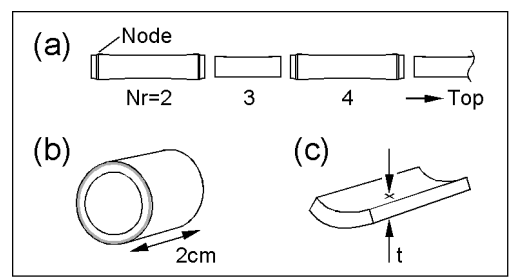

Fig.2. Appearance of cane specimens

Another cane pole was separated into short tubes of 2cm long as shown in Fig.2b. At least 4 tubes were made from each internode. These tubes were then dried in various drying conditions as described below.

Sixteen tubes made from 4 internodes (Nr=10, 15, 19 and 24) were divided into 4 groups (**a** to **d**) and dried at 30°C in an environmental chamber. The groups **a** and **b** were dried at 60%*RH* and 90%*RH* under air-circulation, respectively. The groups **c** and **d** were dried at 90%*RH* without air-circulation, while the group **d** was wrapped with filter paper to retard the drying. After the mass of specimen was reduced by 50% (corresponding to about 25% moisture content), the specimens were dried at 20°C and 65%*RH* for a week.

From 5 internodes (Nr=12, 14, 16, 18 and 20), 20 tubes were made and divided into 4 groups. These groups were dried at 30, 60, 80 and 100°C in an environmental chamber until the weight of specimens were equilibrated. For the drying at 30-80°C, the humidity was kept at 30%*RH*. After the drying, the tubes were cooled and dried at 20°C on $SiO_2$.

Four internodes (Nr=5, 9, 13 and 17) were sectioned into 12 tubes and divided into 3 groups. A part of their surfaces was sealed with silicone grease and aluminum sheet. The tubes were then dried from transverse (I), inner (II) or outer (III) surface at 20°C and 65%*RH* for two weeks with intermittent weighing.

From 4 internodes (Nr=4, 6, 8 and 21), 16 tubes were made and separated into 4 groups. The three groups were dried from I, II or III surface while another group was dried without sealing at 20°C and 65%*RH* for two weeks.

**Surface temperature measurement**

A small piece of 12 mm (L) × 1 mm (R) × 5 mm (T) was made from the inner part of a green cane stem (Nr=2). The sample was hung by a steel frame and dried in an environmental chamber kept at 30°C and 60%*RH*. The weight and surface temperature of the sample were recorded continuously. A pyrometer, Infratherm IN5 (IMPAC Electronic GmbH), was used for the surface temperature measurement. A detailed description of this experimental device can be found in a published paper.[4] The possibilities of sample dimension measurement offered by this device need further investigation and will not be analyzed in the present paper.

**Evaluation of intensity of collapse**

Each cane tube was splinted into 8 to 12 strips. Since the collapse of cane is remarkable in the radial direction,[1] we dealt with the thickness of cane stem as shown in Fig.2c. At the first, these specimens were dried absolutely *in vacuo* on $P_2O_5$ at room temperature and their thickness ($t_1$) was measured at their center part. Next the specimens were humidified at 100%*RH* at room temperature for 1 to 2 weeks and then steamed at 90-96°C for an hour using a cooking steamer. The steamed specimens were cooled in a wet cloth and their thickness ($t_S$) was measured immediately. Finally the specimens were dried again *in vacuo* on $P_2O_5$, and their thickness ($t_2$) was measured. As suggested before[1], the collapse of cane recovers almost completely by the steaming, and the steamed cane shows few re-collapse in the following drying. Therefore, the intensity of collapse remaining in the dry cane was evaluated by the following equation,

$$S_C(\%) = 100\ (t_2 - t_1)/t_S. \qquad (1)$$

## Results and Discussion

### Effect of drying rate

Figure 3 shows the average moisture content (*M*) of cane plotted against the square root of drying duration ($t^{1/2}$). Since the most remarkable collapse of cane occurs above 50%*M*,[1] here we define the drying time ($t_D$) at which the *M* of specimen reaches 50%. The effects of $t_D$ on the intensity of collapse ($S_C$) is shown in Fig.4. Irrespective of internodes, slower drying resulted in larger intensity of collapse. Similar result has been reported for the collapse in wood[2] and bamboo,[3] but no sufficient explanation was given.

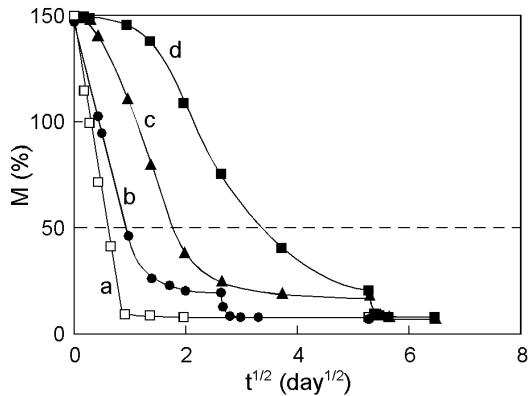

Fig.3. Average moisture content (*M*) of cane specimens (Nr=15) dried at 30°C plotted against the square root of drying duration ($t^{1/2}$). **a**, Dried at 60%*RH* with air circulation; **b**, dried at 90%*RH* with air circulation; **c**, dried at 90%*RH* without air circulation; **d**, wrapped with filter paper and dried at 90%*RH* without air circulation; broken line, a threshold to evaluate the drying time ($t_D$)

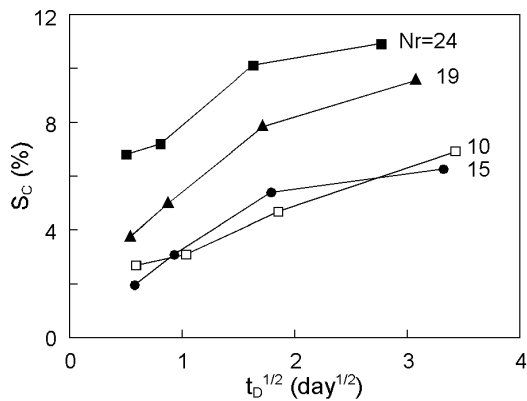

Fig.4. Intensity of collapse ($S_C$) plotted against the square root of drying time ($t_D^{1/2}$). Abbreviations besides plots indicate the drying conditions explained in Fig.3

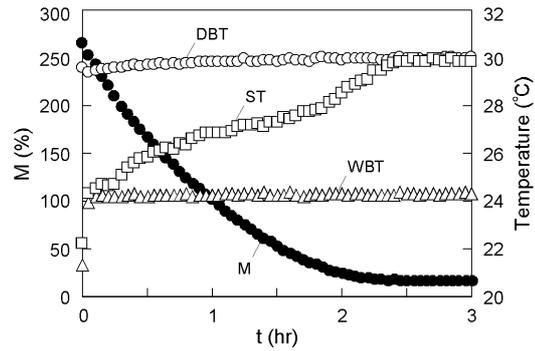

Fig.5. Changes in average moisture content (*M*) and surface temperature (*ST*) of a cane piece with drying at 30°C and 60%*RH*. *DBT*, dry bulb temperature; *WBT*, wet bulb temperature

Keeping in mind that collapse results from the competition between the capillary action (the driving force) and the mechanical behavior of the cell walls (the resisting force), both the process duration and the temperature level have to be involved in the explanation. These two parameters are indeed of utmost importance to the viscoelastic behavior of the cell walls. In addition, the actual sample temperature, rather than the air temperature, should be considered. Figure 5 shows the change in the surface temperature of a wet cane sample during drying at 30°C and 60% *RH*. The surface temperature was very close to the wet bulb temperature (*WBT*) in the early stage of drying, and it gradually approached to the dry bulb temperature (*DBT*) until the fiber saturation point (FSP, *M*≈20%). When a sample is small enough to neglect the internal temperature gradient, the surface temperature can be representative of the sample temperature. Since the collapse of cane occurs above the FSP,[1] the sample temperature relevant to the mechanical phenomenon corresponds to the wet bulb temperature depending on the *RH*. At a given air temperature (*DBT*), both the drying time and the *WBT* increase as the relative humidity increases. These cumulative effects can explain why the drying rate has such an effect on the collapse. In the present case, however, the possible variation in the sample temperature was not so wide, from 24°C (60%*RH*) to 30°C (90%*RH*) where the temperature dependence of collapse was very small as exhibited later. Thus, although the actual sample temperature might strengthen the trend, it does not seem a dominant factor to determine the intensity of collapse, at least in the low temperature range discussed here.

The second interpretation is the shell effect. In general, the cell wall is rigidified with decreasing its moisture content below the FSP. When a sample is rapidly dried, its outer surface is dried and stiffened much faster than the inner part. Consequently that peripheral zone will attain a low moisture content with reduced viscoelastic creep. The existence of such "shell" may prohibit the following internal collapse, whereas it often turns into localized collapse with severe internal checking in the case of wood drying. This interpretation sounds reasonable when we discuss the collapse of cane in its tangential direction. In the tube-like sample used in this study, the tangential collapse must be effectively reduced by the inner surface stiffened by faster drying as well as the hard, waxy and silicated outer surface. However, the collapse of cane is especially remarkable in the radial direction, and the cane tube does not have enough surface to restrict the radial collapse. Thus the shell effect or similar mechanical restriction seems a minor reason for the significant reduction of radial collapse due to rapid drying.

The third explanation is the viscoelastic effect. It is generally accepted that the cell collapse is induced by the liquid tension of free water, except for a few species showing the collapse due to drying stress.[5] If the cane cell wall is an elastic media, no collapse should remain in dry cane because the cells must completely recover their initial shapes after the disappearance of free water *i.e.* the removal of load. However, the collapse of cane actually remains even after the disappearance of free water because the cell wall is viscoelastic and its deformation is not immediately recovered after the removal of load. In addition, a part of strain is fixed by the temporary rearrangement of amorphous molecules, so called drying-set, and it remains unless the materials are well softened by proper hygro-thermal treatment. Thus the collapse of cane and its recovery by steaming are understood as the viscoelastic yield of cell wall and the release of drying-set, respectively. When we deal with the collapse as a viscoelastic phenomenon, it is quite natural that longer loading due to slower drying results in more remarkable or frequent yield of cells, and also, it expands the duration required for the recovery of collapse. This explanation is also valid for the effect of drying temperature described later.

**Effect of drying temperature**

Figure 6 exhibits the effect of drying temperature on the intensity of collapse. The intensity of collapse increased with increasing the drying temperature irrespective of the internodes. As described above, faster drying results in less collapse at around room temperature. However, it has been suggested that the dynamic Young's modulus of wet cane drops at its softening point, about 90°C.[6] This fact indicates that the wet cell wall yields easier at higher temperature because of its hygro-thermal softening. In addition, it is considered that the faster drying does not allow the recovery of collapse after the disappearance of free water, while it effectively fixes the remaining collapse in terms of drying-set. These may be the reason for greater collapse at high temperature.

Although many factors must be involved, the viscoelastic effect seems the most important factor to determine the intensity of collapse. It can explain the effects of drying rate and heating temperature at the same time, that is, time-temperature dependent phenomenon. For more detailed discussion, the static viscoelastic behavior of cane should be clarified in the future.

drying and then leveled off. From the linear correlation of $\Delta mA^{-1}$ vs. $t^{1/2}$ in the range from 2hr to 24hr, the drying rate ($\Delta mA^{-1} t^{-1/2}$) was evaluated. The drying rate is plotted against the node number (Nr) in Figure 8. Irrespective of internodes, the drying rate of transverse section is twice larger than that of the inner surface, and about 7 times larger than that of the outer surface. The rapid drying from the transverse surface was attributable to the large vessels in vascular bundles, and the especially slow drying from the outer surface may be due to its dense and silicated structure.

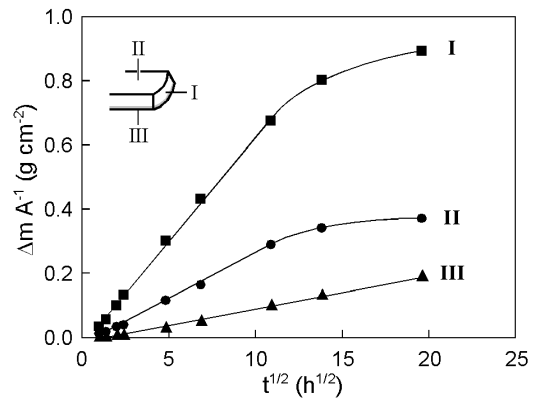

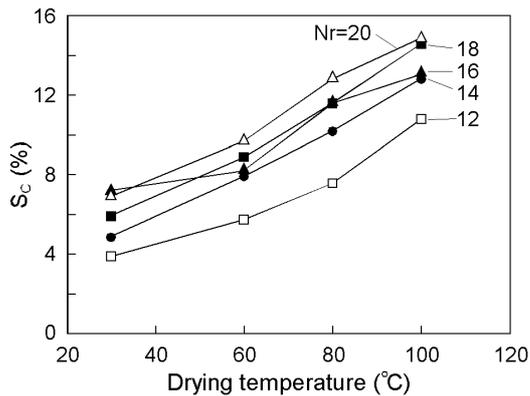

Fig.6. Effect of drying temperature on the intensity of collapse ($S_C$)

Fig.7. Reduction in mass ($\Delta mA^{-1}$) of green cane specimens (Nr=5) due to drying from transverse (I), inner (II) or outer (III) surface with the elapse of time ($t^{1/2}$). $\Delta m$, reduction in mass; $A$, area of drying surface

**Effect of drying surface**

The cane stem has waxy outer layer where silica and other inorganic substances are condensed.[7] Such layer was thought to retard the drying and to affect the collapse. Figure 7 shows the reduction in mass ($\Delta mA^{-1}$) of green cane specimens due to drying from transverse (I), inner (II) or outer (III) surfaces with the square root of drying duration ($t^{1/2}$). The reduction in mass ($\Delta m$) was normalized by the area ($A$) of open surface. The $\Delta mA^{-1}$ value increased linearly with the $t^{1/2}$ in the beginning of

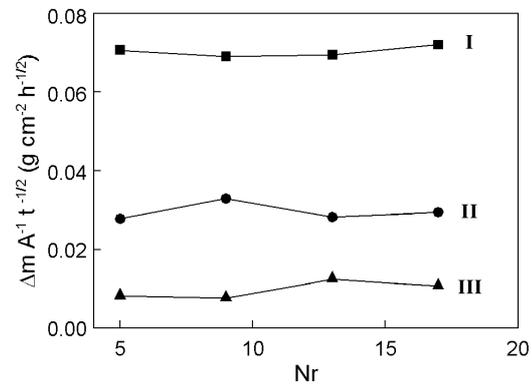

Fig.8. Drying rate ($\Delta mA^{-1} t^{-1/2}$) at different surface of cane specimens plotted against the position of internode (Nr). For abbreviations, see Fig.7

Figure 9 shows the reduction in $M$ due to drying from different surfaces, and the $S_C$ values of the specimens are shown in Fig. 10. The drying from the outer surface caused the most remarkable collapse probably because of slower drying, whereas faster drying from inner surface resulted in the least collapse. Interestingly, the drying from transverse section gave relatively large $S_C$ value. It should be recalled that the cane specimen tested was only 2cm long, and steep moisture gradient can hardly be formed along the fiber direction. Furthermore, serious collapse was always observed in the middle layer where the parenchyma cells were less frequent than the inner layer. A possible interpretation is that the differential drying of different tissues induced the collapse. The cane mainly consists of vascular bundles and parenchyma cells. As the vascular bundle has large continuous vessels, it may dry much faster than the parenchyma cells. In this case, steep moisture gradient can be formed between those two tissues to cause the collapse of parenchyma cells. Otherwise the shrinkage of thick cell wall in bundle sheaths might be a trigger for the collapse of surrounding parenchyma cells.

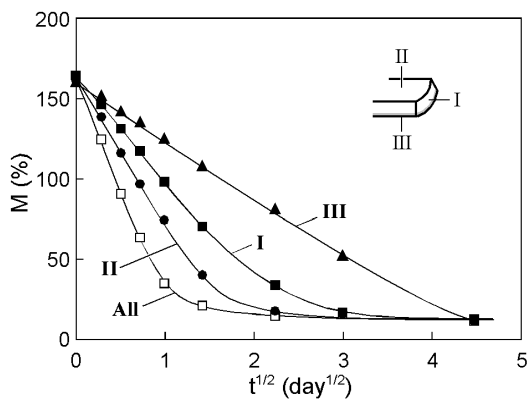

Fig.9. Changes in the average moisture content ($M$) of cane specimens (Nr=8) with the square root of drying duration ($t^{1/2}$). All, Dried from all surfaces (unsealed); I, dried from transverse surface; II, dried from inner surface; III, dried from outer surface

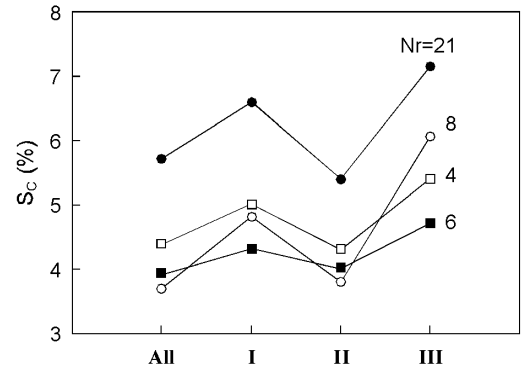

Fig.10. Intensity of collapse ($S_C$) of cane specimens dried from different surfaces. For abbreviations, see Fig.9

**Effect of node**

Since the presence of node retards the drying from the transverse and inner surfaces of cane, the cane tube having node dried much slower than that without node. Figure 11 shows the effect of node on the intensity of collapse. In the presence of node, the intensity of collapse increased above 8$^{th}$ node probably due to the retardation of drying.

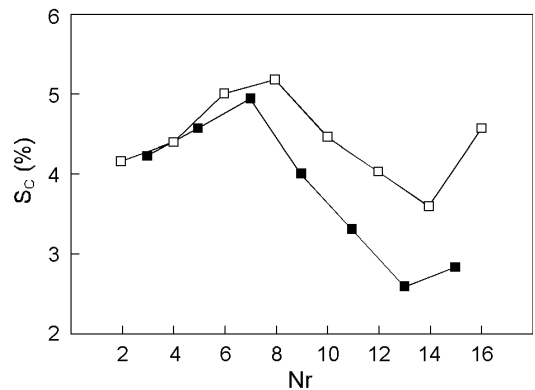

Fig.11. Intensity of collapse ($S_C$) for cane specimens plotted against the node number (Nr). *Open plots*, with node; *closed plots*, without node

For making the reeds of woodwind instruments, the cane poles are usually dried in open air without being separated. However, all experimental results indicate that such a very slow drying leads to greater intensity of collapse, especially when the cane is dried from its outer surface. In addition, slower drying in highly humid condition (above 90%*RH*) sometimes results in serious stain due to fungi. Thus it is advisable to remove the node and to dry faster at lower temperature, for better quality of final products.

**Conclusion**

Cane specimens were dried in various conditions and the intensity of collapse was evaluated. The results are concluded as follows:

1) The intensity of collapse was increased by slower drying. It was considered that higher sample temperature and longer loading of liquid tension due to slower drying induced more remarkable collapse, while rapid drying stiffened the surface to restrict the following collapse.

2) The intensity of collapse increased with increasing drying temperature. The effect of heating was explained by the easier yield of the cell wall being thermally softened, and the rapid drying-set of collapsed cells restricting their recovery.

3) Drying from waxy outer surface or the presence of node resulted in greater intensity of collapse probably due to the retardation of drying. Thus it is advisable to remove the node and to dry faster at lower temperature, for drying cane with less collapse.

**Acknowledgments**

The authors are sincerely grateful to Franco Guccini, Marca Reed Inc. for kindly providing cane samples.